\documentclass[12pt]{iopart}
\pdfoutput=1
\expandafter\let\csname equation*\endcsname\relax
\expandafter\let\csname endequation*\endcsname\relax
\usepackage[numbers,sort&compress]{natbib} 
\usepackage[pdftex]{graphicx,color}
\usepackage{amsmath,amsfonts,amssymb,wasysym,graphicx,capt-of,ifthen,calc}
\usepackage{enumerate,float,url,color,subfigure}
\usepackage{amsmath}
\usepackage{graphics}  
\usepackage{psfrag} 
\usepackage[latin1]{inputenc}
\usepackage{color}
\usepackage{rotating}
\usepackage{url}
\usepackage{hyperref}
\def\etl{$et ~al.$~}

\begin{document}

\title{Energy fluctuations in one dimensional Zhang sandpile model}
\author{Naveen Kumar}
\address{Department of Physics \& Astronomical Sciences, Central University of Jammu, Samba 181 143, India}

\author{Suram Singh}
\address{Department of Physics \& Astronomical Sciences, Central University of Jammu, Samba 181 143, India}

\author{Avinash Chand Yadav}
\address{Department of Physics, Institute of Science,  Banaras Hindu University, Varanasi 221 005, India}

\begin{abstract}
We consider the Zhang sandpile model in one-dimension (1D) with locally conservative (or dissipative) dynamics and examine its total energy fluctuations at the external drive time scale. 
The bulk-driven system leads to Lorentzian spectra, with a cutoff time $T$ growing linearly with the system size $L$.  
The fluctuations show $1/f^{\alpha}$ behavior with $\alpha \sim 1$ for the boundary drive, and the cutoff time varies non-linearly.  For conservative local dynamics, the cutoff time shows a power-law growth $T \sim L^{\lambda}$ that differs from an exponential form $ \sim \exp(\mu L)$ observed for the nonconservative case. We suggest that the local dissipation is not a necessary ingredient of the system in 1D to get the $1/f$ noise, and the cutoff time can reveal the distinct nature of the local dynamics. We also discuss the energy fluctuations for locally nonconservative dynamics with random dissipation.  
\end{abstract}

\maketitle

\section{Introduction}
Astonishingly diverse systems exhibit temporal fluctuations characterized by a low-frequency $1/f^{\alpha}$-type power spectral density (PSD), with the spectral exponent $\alpha$ lying between 0 and 2~\cite{Dutta_1981, Weissman_1988, Eliazara_2009, Erland_2011, Yadav_2013, Yadav_2017, Sposini_2020}. Typical examples range from voltage fluctuations across a resistor to rhythm variations in music~\cite{Menon_2011}. While a linear operation, integration of white noise ($\alpha = 0$) yields Brownian noise ($\alpha = 2$), a non-integer value of the spectral exponent reflects the nonlinear nature of the underlying process. One aspect that led to a significant amount of research is the origin of $1/f$ noise. Previous studies suggest the existence of a few underlying mechanisms. Self-organized criticality (SOC) and a superposition of Lorentzian spectra with power-law distributed relaxation time are the most common. SOC introduced by Bak, Tang, and Wiesenfeld (BTW)~\cite{Bak_1987, Bak_1996, Gros_2014} implies that a class of non-equilibrium systems, when driven slowly, respond instantaneously as avalanches that follow power-law distribution in the critical state. The criticality emerges spontaneously because of self-organization. Although BTW proposed a cellular automaton (termed as BTW sandpile model) to explain the $1/f$ noise by relating scaling features of avalanches with the spectral property, a few subsequent studies~\cite{Jensen_1989, Kertesz_1990} argued the noise to be $1/f^2$-type, supported with simulation studies for smaller system size. However, based on a careful scaling analysis, later studies established that the BTW sandpile model indeed shows $1/f$ behavior~\cite{Laurson_2005}.

Several variants of the BTW sandpile model also show $1/f$ noise. It is also interesting to mention that the studies of $1/f$ noise in sandpiles have examined signals monitored on different time scales. One example is the avalanche activity signal, the number of toppling during a parallel update as a function of time (the fast time scale). It reflects the dynamics within avalanches~\cite{Travesset_2002}. Other examples include the total energy (or mass) fluctuations observed at the external drive time scale (the slow time scale), capturing the time correlations between avalanches. The total mass fluctuations at the drive time scale of a directed sandpile model with discrete state and locally conservative dynamics, studied by Maslov~\emph{et al.}~\cite{Maslov_1999} and later exactly solved in Ref.~\cite{Yadav_2012}, display a clear $1/f$ noise with $\alpha \approx 1$. A class of boundary driven SOC systems (train model~\cite{train_model, Davidsen_2002} and Oslo sandpile~\cite{Christensen_1996}) displays scaling properties for space-time correlations characterized by both critical avalanches and $1/f^{\alpha}$ noise. Christensen~\etl\cite{Christensen_1992} showed $1/f$ noise in a SOC system with locally non-conservative dynamics. For deterministic lattice gases~\cite{jensen1, giometto}, the density-fluctuations show $1/f$ power-spectrum.

The spectral exponent may differ from 1 and depend on the dimension. Also, SOC cannot be a necessary condition since  $1/f$ noise and power-law spatial correlation (critical avalanches) do not always coexist. Motivated by these ideas, DeLosRios and Zhang~\cite{Zhang_1999} studied a continuous state sandpile (Zhang sandpile) with boundary drive and locally nonconservative dynamics. The system can explain $1/f$ noise, with spectral exponent  $ \approx 1$ (ideal $1/f$ noise) and independent of dimension (hyper-universal). The model has two salient features. (i) The energy flow has a preferred direction, driving the system at one boundary. (ii) In local redistribution, some energy is lost (nonconservative). Locally nonconservative feature destroys SOC, and the power-law distribution for avalanche sizes does not sustain. However, some features survive, like self-organization, as the probability density function of energy shows a few peaks with finite spread. The directed Zhang sandpile model with nonconservative property shows $1/f$ noise for the fluctuations in the total energy. Its explanation is straightforward to follow as a superposition of locally independent Lorentzian spectra with a cutoff time growing exponentially with distance from the driving end.

We revisit the Zhang sandpile model in 1D with locally conservative (or dissipative) dynamics. The quantity of interest is the fluctuations in the total energy at the external drive time scale. A trivial behavior is Lorentzian spectra (for bulk drive) with a cutoff time $T$ that grows linearly with the system size $L$.  
For the boundary drive, the fluctuations in the total energy show $1/f^{\alpha}$ behavior with $\alpha \sim 1$, but the cutoff time varies non-linearly. Interestingly, when the local dynamics remain conservative, the behavior of cutoff time shows a power-law growth $T \sim L^{\lambda}$, different from a known exponential form $ \sim \exp(\mu L)$ observed with nonconservative dynamics. The local dissipation is not an essential element for holding the $1/f$ noise in the system. It also implies that the behavior of cutoff time as a function of system size can reveal the nature of system dynamics.

The organization of the paper is as follows. We recall the model definition in Section~\ref{sec_ii}. We outline the numerical methods in Section~\ref{sec_iii} and present the analytical and numerical results for the energy fluctuations with locally conservative and nonconservative dynamics in Section~\ref{sec_iv} and ~\ref{sec_v}, respectively. Finally, Section~\ref{sec_vi} provides a summary and discussion, followed by an appendix showing additional results for the bulk drive.

\section{Zhang sandpile model}{\label{sec_ii}}
We consider the Zhang sandpile model~\cite{Zhang_1999} on a 1D lattice with $L$ sites. It is a continuous state variant of the BTW sandpile. The boundaries are open at both ends and allow a loss of excess flow. Generalization to a higher dimension is straightforward. At each site $i$, assign a continuous state variable (energy) $z(i)$. The state is stable if $0\le z(i)<z_{0}$, where $z_{0}$ is a threshold. We initialize the system by assigning $z(i)$ to be a random variable, with a uniform distribution between 0 and $z_{0}$.
The dynamics comprise two elementary steps. (i) At each time step $t$,  an uncorrelated noise drives the system at one boundary 
\begin{equation}
z(1) = z(1) +\delta,\nonumber
\end{equation}
where $\delta$ is a random variable with uniform distribution in an interval $[0, \delta_m]$. (ii) As a result of driving, a site may become unstable $z(i)\ge z_{0}$. If this happens, then relaxation occurs by the following redistribution rules $z(i) \to 0$ and
\begin{eqnarray}
z(i\pm 1) \to  z(i\pm 1) + (1-a)\frac{z(i)}{2},
\label{eq_rd_rule}
\end{eqnarray}
where $a\in [0, 1)$ is a dissipation parameter. 
The presence of local dissipation makes the redistribution process locally nonconservative.

An unstable site relaxes by transferring its energy to the nearest neighbors, and the neighbor site(s) may become unstable. The active site(s) may further trigger, and the relaxation continues until these become stable. The cascade event is an avalanche, and the next drive applies when the avalanche activity is over.  
The addition of energy makes the process grow randomly in time, and occasional dissipation from the boundary brings it down. We can also find models showing such behavior in a class of processes characterized by unidirectional random growth with reset events~\cite{Neda_2018}. Ultimately, the total energy evolves in a fluctuating manner in response to such a drive. Our interest is in the total energy of the system, expressed as the sum of local energies  
 \begin{equation}
 \xi(t) = \sum_{x=1}^{L}z(x,t).\nonumber
\end{equation}

\begin{figure}[t]
  \centering
  \scalebox{0.65}{\includegraphics{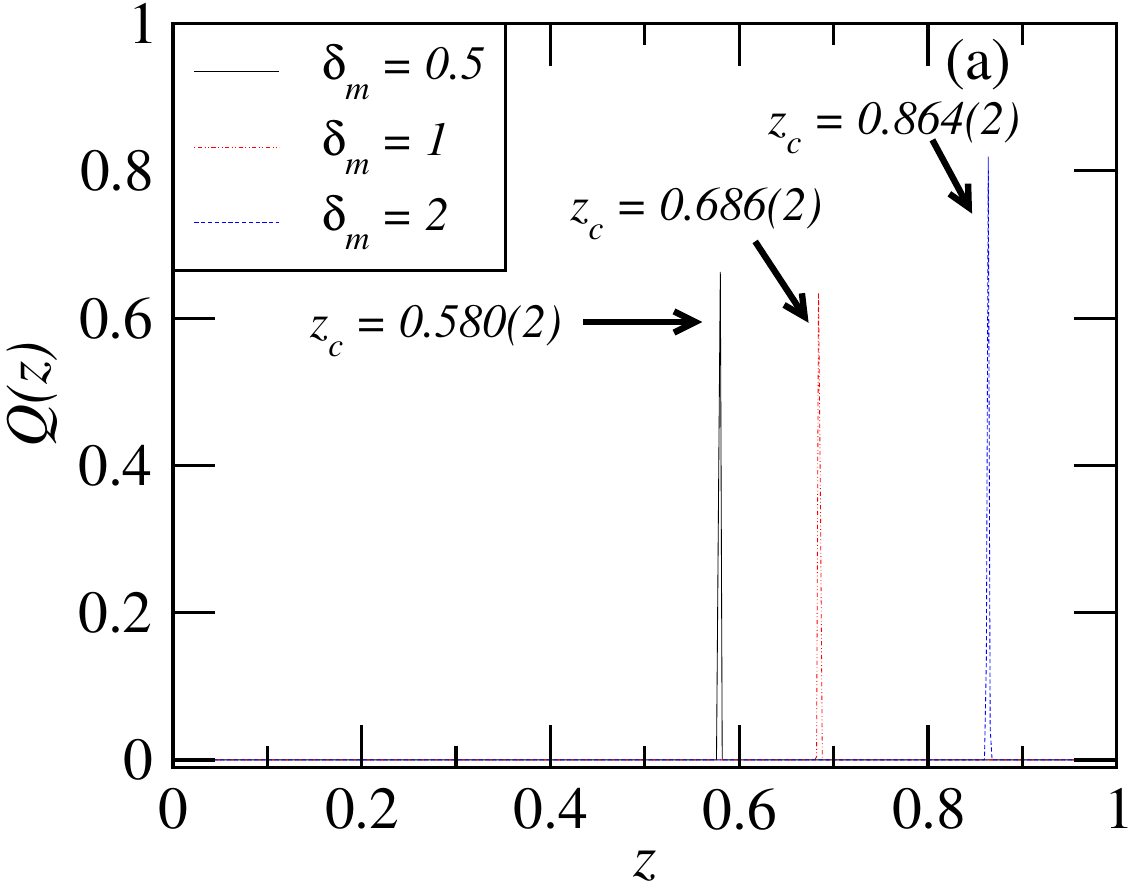}}
   \scalebox{0.62}{\includegraphics{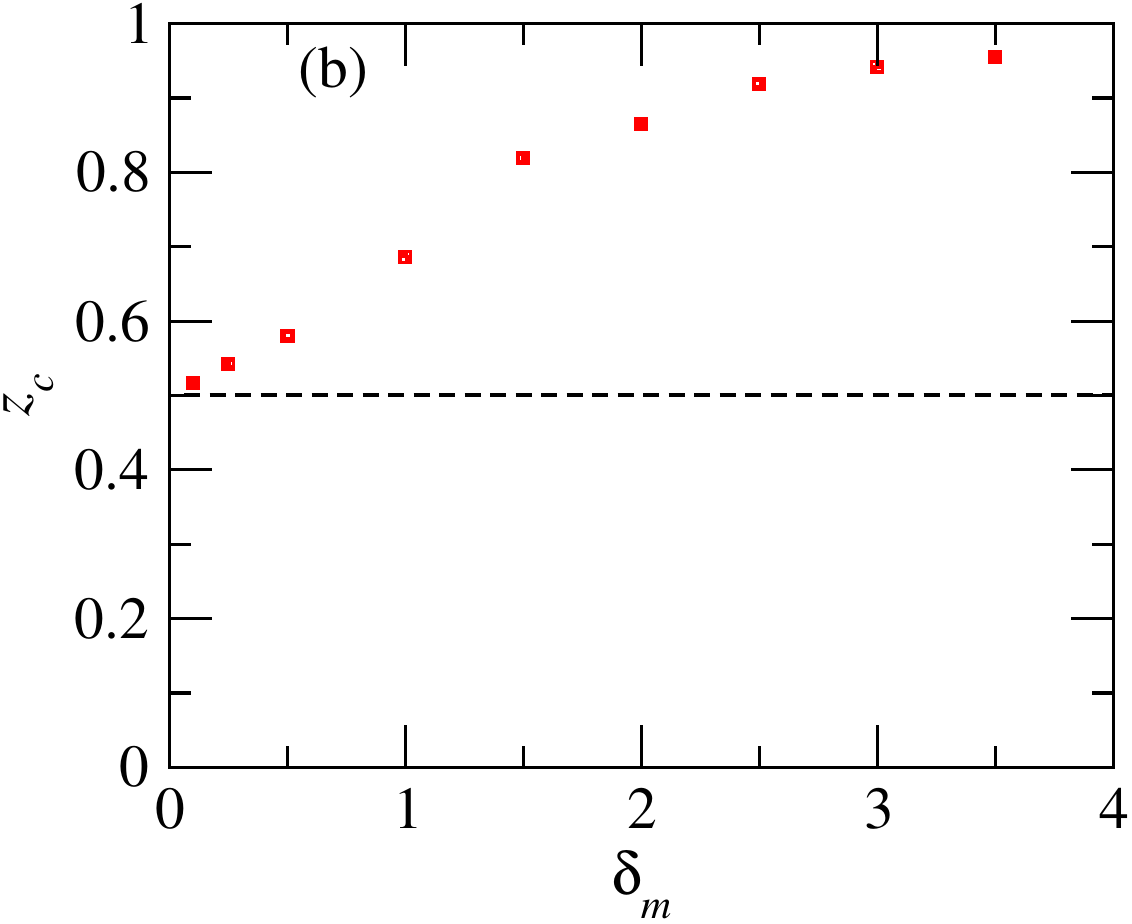}}
  \caption{(a) The probability density function of the energy, $Q(z)$. The three curves (different peaks with finite spread) correspond to different values of the parameter $\delta_m$. A trivial peak is at $z = 0$ with $Q(z)>0$, but it is not visible as the probability value is relatively small. The nontrivial peak happens at $z_c$, the average energy per site. We obtain the distribution by collecting the energies of a site with $10^8$ samples on the external drive time scale. Here, $a = 0$, $z_0 = 1$, and $L=2^{9}$. The resolution of energy is $\Delta z = 0.002$. We store the data after discarding the transients. An initial configuration with $z(i) \to z_c$ reduces the transient effect. (b) The variation of $z_c$ with $\delta_m$. The dashed line marks $z_c = 0.5$.}
   \label{fig_qz_1d}
\end{figure}

For $a = 0$, a power-law distribution for size or duration characterizes the critical avalanches. The other striking features of the system are the following. 
In the critical state, self-organization occurs, where the average energy attains a critical value $z_c$ (a fixed value in the limit of large system size). Also, the probability density of energy, $Q(z)$, shows a few peaks with finite spread, irrespective of the initial uniform distribution~\cite{Zhang_1989}. The number of peaks is 2D on a D-dimensional hyper-cubic lattice with nearest-neighbor interaction, and these appear at the energy values 0, $z_0$/2D, 2$z_0$/2D, $\dots$, (2D-1)$z_0$/2D.

In 1D, Fig.~\ref{fig_qz_1d} (a) shows the probability density of the energy, $Q(z)$, for different values of the parameter $\delta_m$. Here, the most probable value of the local energy coincides with $z_c$.  With $z_0 = 1$, the non-trivial peak occurs at $z_c = 1/2$ in one dimension, if $\delta_m \ll z_0$. 
Increasing $\delta_m$ only shifts $z_c$ towards 1 [cf. Fig.~\ref{fig_qz_1d} (b)]. The variation of $z_c$ with $\delta_m$ is nonlinear with an increasing trend and eventually saturating to 1. For $0< a < 1$, the power-law spatial correlation does not persist. It destroys SOC as avalanches are no more critical, but the probability density function of energy retains the peaks~\cite{Zhang_1999}. Numerically, we also observed that the dissipation parameter does not change the value of energy where the peak occurs.

\section{Numerical Methods}{\label{sec_iii}}
We examine the energy fluctuations in 1D Zhang sandpile model by computing their power spectra. In this section, we mention the numerical methods that we employ. Without loss of generality, we can set the threshold state $z_{0}=1$ in the simulation. If $\delta_m\gg z_0$, the system always yields an avalanche for each drive. However, if $\delta_m\ll z_0$, an addition of input noise may not trigger the system occasionally. The short time quiet behavior gets reflected as a flat PSD at higher frequencies because of the uncorrelated nature of the input noise~\cite{Zhang_1999}. As $z(x,t) \in [0,1)$, it is a stationary process in the long time limit. Similarly, the total energy, a finite sum of local energies is also a stationary process. We compute the PSD as
$S(f) = \lim_{N\to \infty}\langle |\hat{Y}(f;N)|^2\rangle/N$, 
where $\hat{Y}(f;N)$ is a Fourier-series transform of a noisy signal $Y(t)$ in a time-interval $t\in [0, N]$. The symbol $\langle \cdot \rangle$ denotes the ensemble average. 
We use Monte Carlo methods to simulate the process and compute the power spectrum employing the standard fast Fourier transform algorithm. We perform the ensemble average with $M$ different realizations of the time series, each of length $N$. The PSD has a cutoff in frequency $1/T$, where $T$ is the cutoff time and the long time limit $N\gg T$ applies.

\begin{figure}[b]
  \centering
  \scalebox{0.62}{\includegraphics{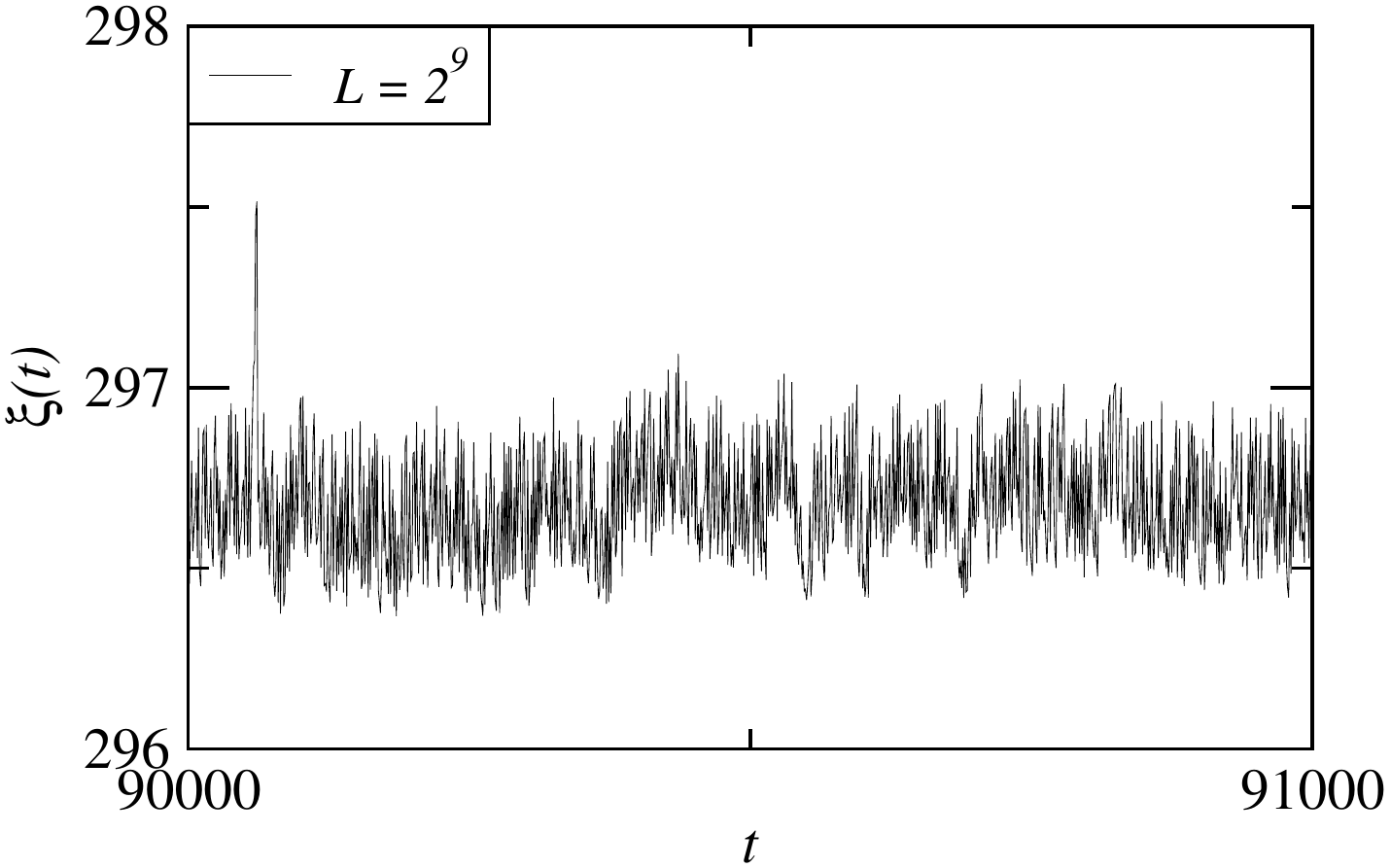}}
  \caption{A typical signal for the total energy $\xi(t)$ with the system size $L = 2^9$.}
   \label{fig_sg_1d_l}
\end{figure}

\begin{figure}[t]
  \centering
  \scalebox{0.7}{\includegraphics{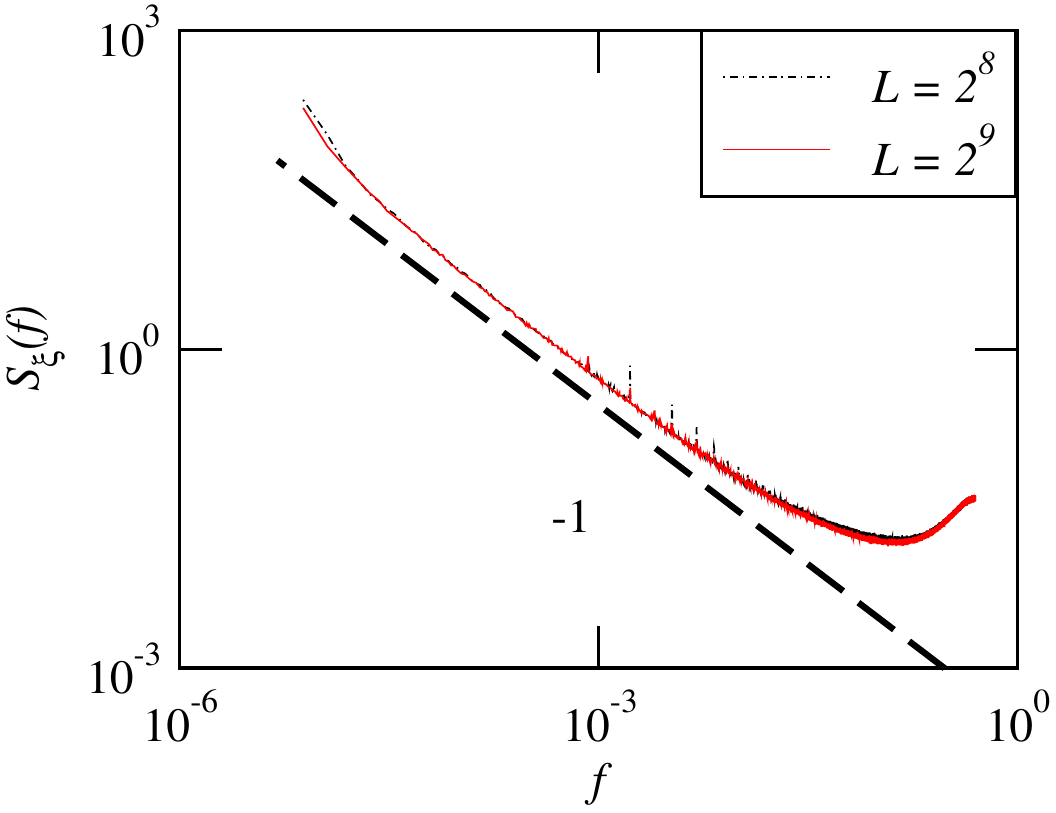}}
  \caption{The power spectrum $S_{\xi}(f)$ for  the fluctuations in the total energy, $\xi(t)$, with two different values of the system size $L = 2^8$ and  $2^9$.  The dashed line has a slope -1. The signal length is $N  = 2^{18}$, and ensemble averaging is performed over $M = 2 \times 10^3$ different realizations.}
   \label{fig_ps_1d_l}
\end{figure}

\begin{figure}[t]
  \centering
   \scalebox{0.72}{\includegraphics{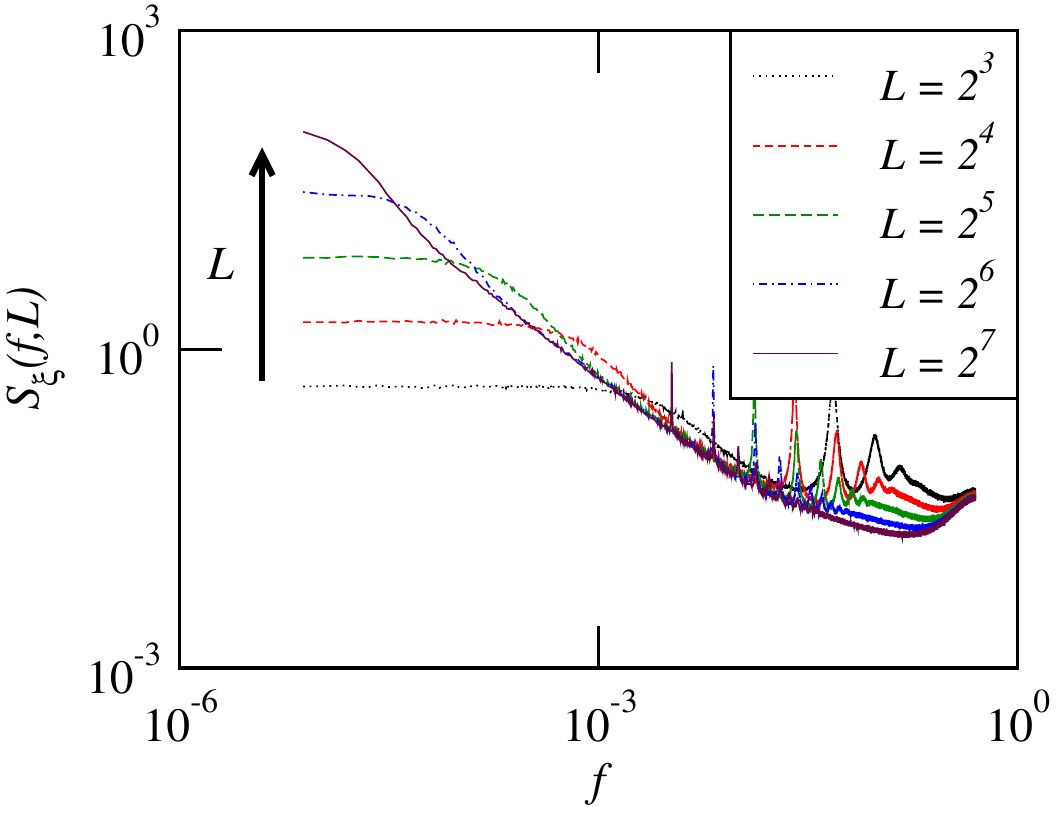}}
  \caption{The power spectra for  $\xi(t)$ show a cutoff as a function of system size $L$.  }
  \label{fig_ps_1d_T}
\end{figure}

\begin{figure}[t]
  \centering
   \scalebox{0.72}{\includegraphics{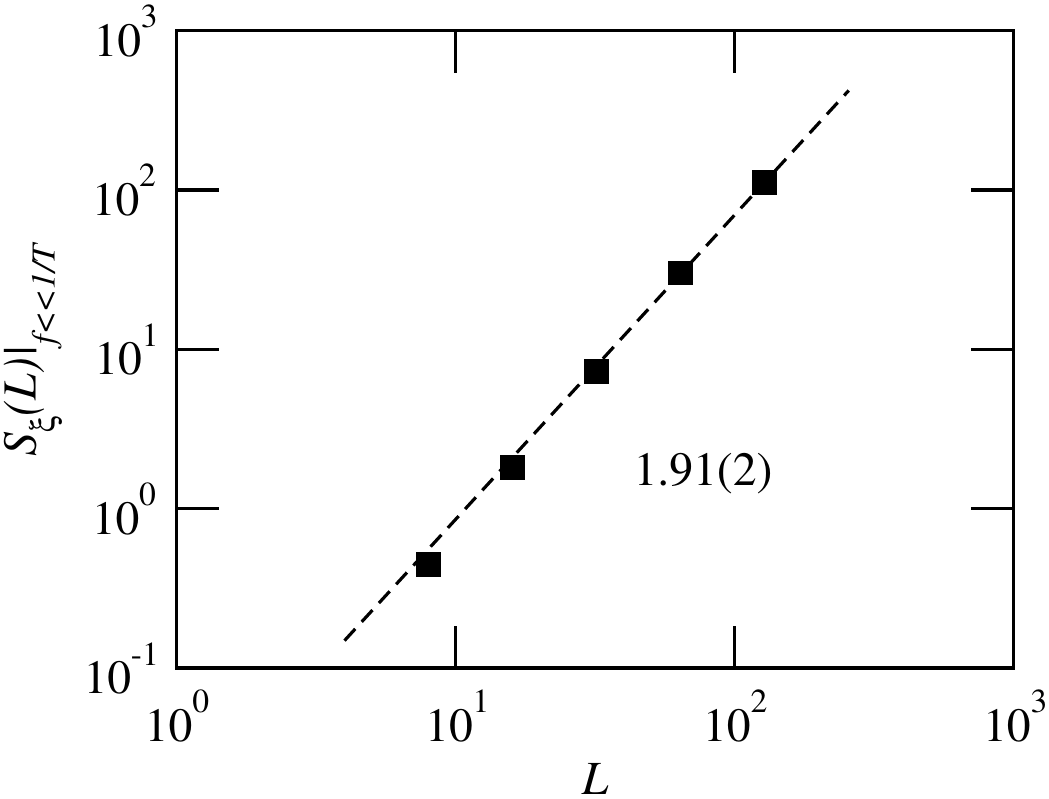}}
  \caption{ The power in low frequency component as a function of system size: $ S_{\xi}(L)|_{f\ll 1/T} \sim T\sim L^{\lambda}$ with $\lambda  = 1.91(2)$. } 
  \label{fig_ps_1d_T_1}
\end{figure}

\begin{figure}[t]
  \centering
   \scalebox{0.72}{\includegraphics{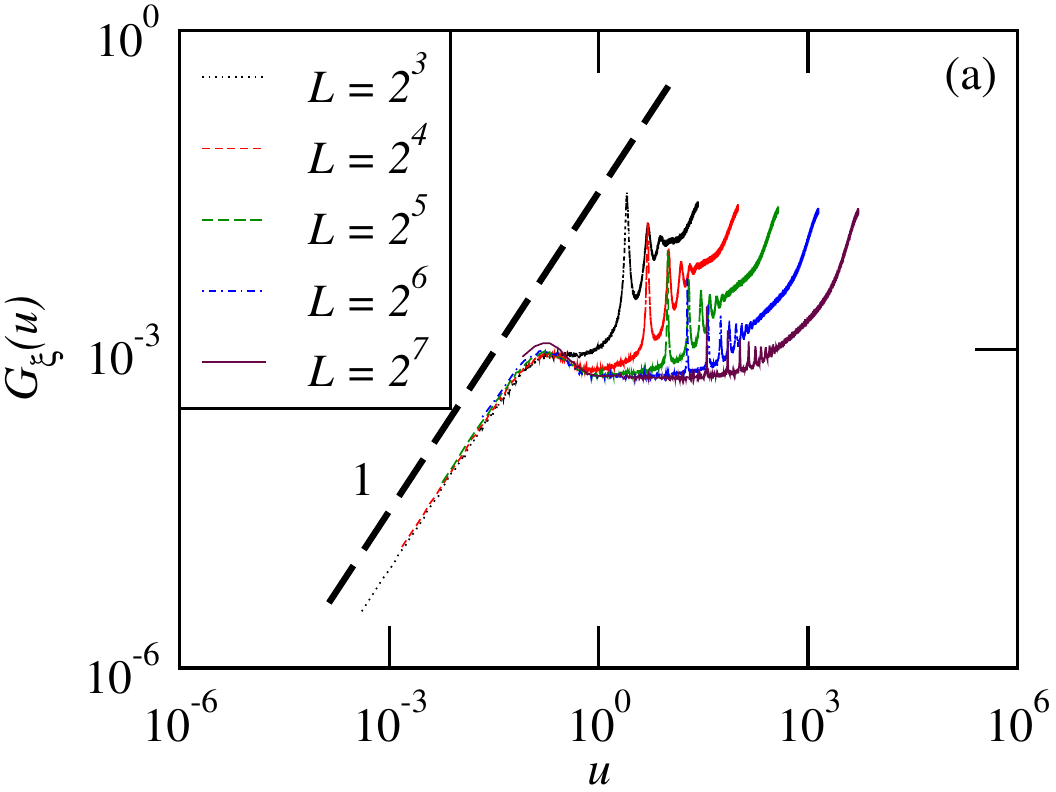}}
   \scalebox{0.72}{\includegraphics{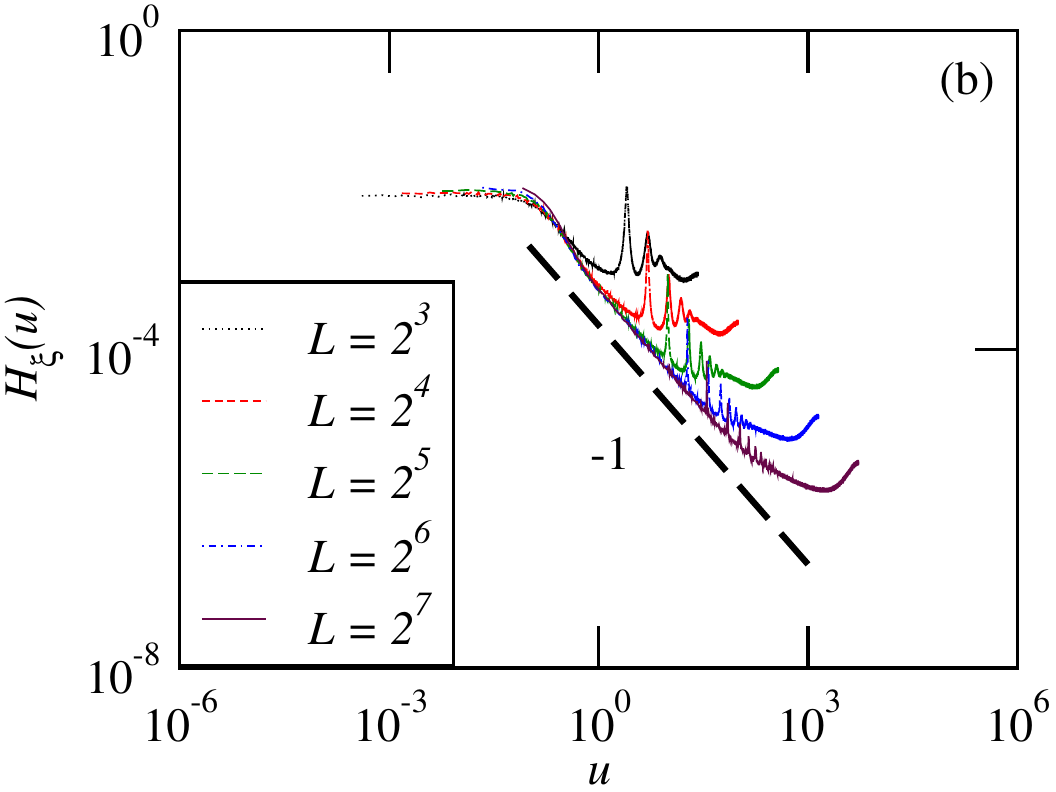}}
  \caption{The scaling functions. (a) $G_{\xi}(u) \sim fS_{\xi}(f,T)$ and (b) $H_{\xi}(u) \sim T^{-1}S_{\xi}(f,T)$ with $u = fT$. }
  \label{fig_1d_ps_dcc}
\end{figure}

 \section{Fluctuations in locally conservative dynamics}\label{sec_iv} 
For locally conservative dynamics, we use $\delta_m = 0.5$ in simulations. We show a typical time series for the fluctuations in the total energy, $\xi(t)$, in Fig.~\ref{fig_sg_1d_l}. Figure~\ref{fig_ps_1d_l} shows power spectra for $\xi(t)$ with two different system sizes, revealing the existence of a clear $1/f$ noise and the noisy peaks reduce on increasing the system size. The spectrum also shows the existence of a cutoff in frequency (a function of the system size). The power is independent of frequency below the cutoff but increases by increasing the cutoff time $T$. When the system size $L$ is doubled, the power on the double log scale increases by a constant amount, suggesting a power-law behavior for the cutoff time $T\sim L^{\lambda}$ [cf. Fig.~\ref{fig_ps_1d_T}]. Then we can write  
 \begin{equation}
S_{\xi}(f,T) = \begin{cases} AT, ~~~~~~{\rm for}~~f \ll 1/T, \\ A/f , ~~~~~{\rm for}~~1/T \ll f \ll 1/2. \end{cases}
\label{a0_psd_eq1}
\end{equation}

Since $S_{\xi}(f,T)$ is a homogenous function of its arguments [cf. Eq.~(\ref{a0_psd_eq1})],  we can apply the scaling method~\cite{Naveen_2021, Yadav_2022} to get
 \begin{equation}
S_{\xi}(f,T) = A\frac{1}{f}G_{\xi}(u) = AT H_{\xi}(u),
\label{a0_psd_eq2}
\end{equation}
where $u=fT$. In Eq.~(\ref{a0_psd_eq2}), $G_{\xi}$ and $H_{\xi}$ are the scaling functions
 \begin{subequations}
 \begin{align}
G_{\xi}(u) \sim \begin{cases} u, ~~~~~~{\rm for}~~u \ll 1, \\ 1, ~~~~~~{\rm for}~~u\gg 1,\end{cases}
\label{a0_psd_eq3_g}
\end{align}
and
\begin{align}
H_{\xi}(u) \sim \begin{cases} 1, ~~~~~~~~~{\rm for}~~u \ll 1, \\ 1/u, ~~~~~~{\rm for}~~u\gg 1.\end{cases}
\label{a0_psd_eq3_h}
\end{align}
\end{subequations}

To get the scaling functions numerically, the exponent $\lambda$ needs to be determined. Notice that the power in low-frequency components varies as [cf. Fig.~\ref{fig_ps_1d_T_1}]
\begin{equation}
S_{\xi}(L)|_{f\ll 1/T} \sim T \sim L^{\lambda}.
\label{a0_psd_eq4}
\end{equation} 
 One can numerically estimate the unknown exponent by plotting the power in low-frequency components as a function of the system size [cf. Eq.~(\ref{a0_psd_eq4})]. Using best fit, we find $\lambda = 1.91(2)$. Figure~\ref{fig_1d_ps_dcc} shows the data collapse curves, consistent with Eqs.~(\ref{a0_psd_eq3_g}) and~(\ref{a0_psd_eq3_h}).
The numerical results suggest that a clean $1/f$ noise emerges in the boundary-driven 1D Zhang model even when the local dynamics are conservative. It is worthy to note that not removing transients leads to the spectrum $1/f^2$ behavior for large system sizes.

\begin{figure}[t]
  \centering
  \scalebox{0.72}{\includegraphics{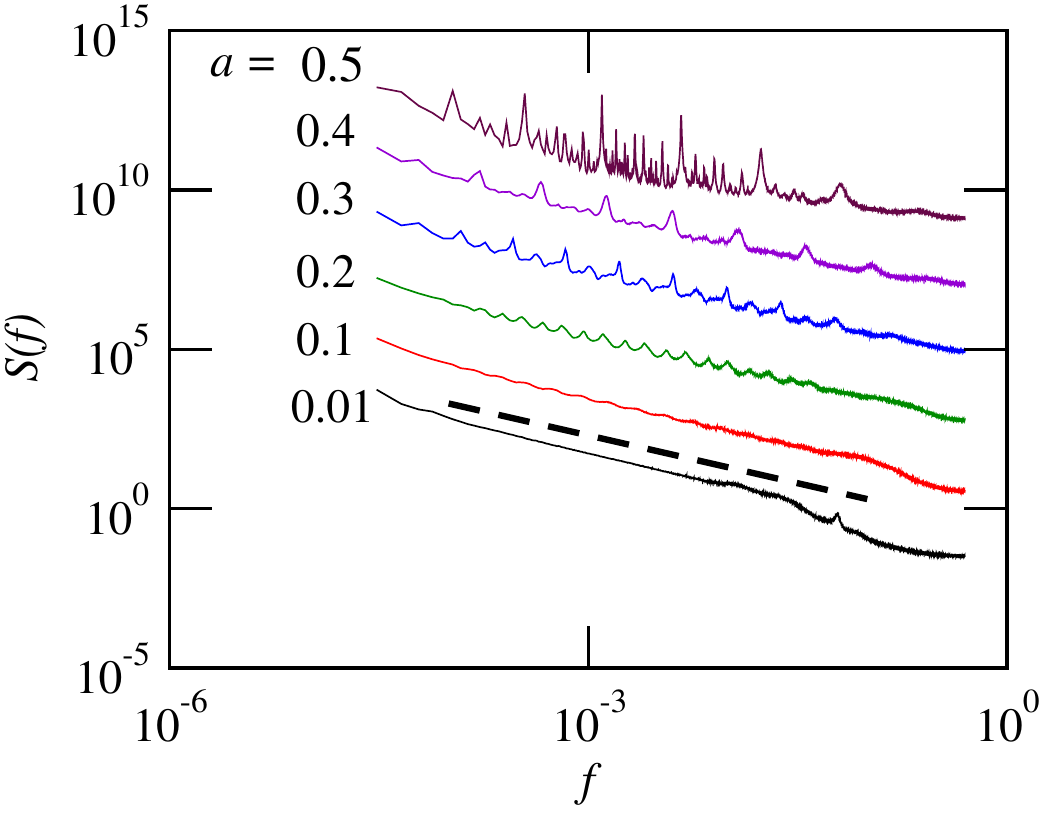}}
  \caption{Power spectra for the fluctuations in the total energy  $\xi(t)$ for different values of constant dissipation parameter $a$ with $N = 2^{16}$ and $M = 10^4$.  A dashed line with slope -1 is drawn for comparison. Each curve is shifted by two decades for clarity.}
  \label{fig_ps_a}
\end{figure}

\section{Fluctuations in locally nonconservative dynamics}\label{sec_v} 
As studied in Ref.~\cite{Zhang_1999}, one way to understand locally nonconservative dynamics is to introduce a constant local dissipation with strength $a$ in the redistribution rules [cf. Eq.~(\ref{eq_rd_rule})]. For local dissipative dynamics, we use $\delta_m = 2$. In Fig.~\ref{fig_ps_a}, we show the power spectra for $\xi(t)$ with different values of the constant local dissipation parameter $a$. We find that on increasing $a$ (beyond $ \sim 0.2$), the $1/f$ spectrum gets dominated by peaks or it does not remain smooth. Further, if we assume $a(x,t)$ to be a random variable, then it captures evolving heterogeneity of the medium. For simplicity, we consider $a$ a random variable with uniform distribution in the unit interval. As shown below, the spectrum remains smooth.

\begin{figure}[t]
  \centering
  \scalebox{0.62}{\includegraphics{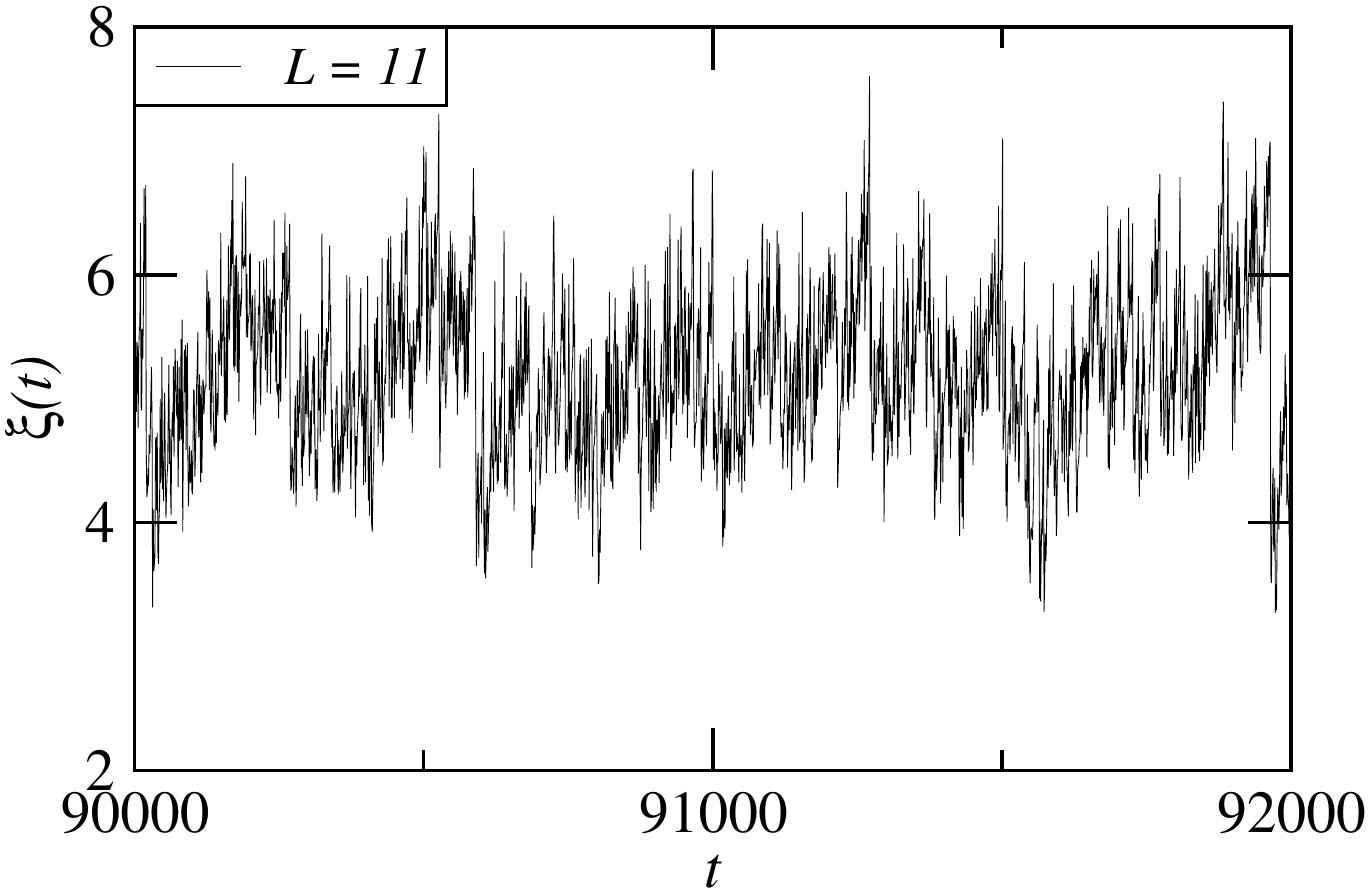}}
  \caption{The total energy $\xi(t)$ as a function of time with the system size $L = 11$. $a(x,t)$ is a random number with a uniform distribution in the unit interval. }
   \label{fig_sg_xi}
\end{figure}

\begin{figure}[t]
  \centering
  \scalebox{0.72}{\includegraphics{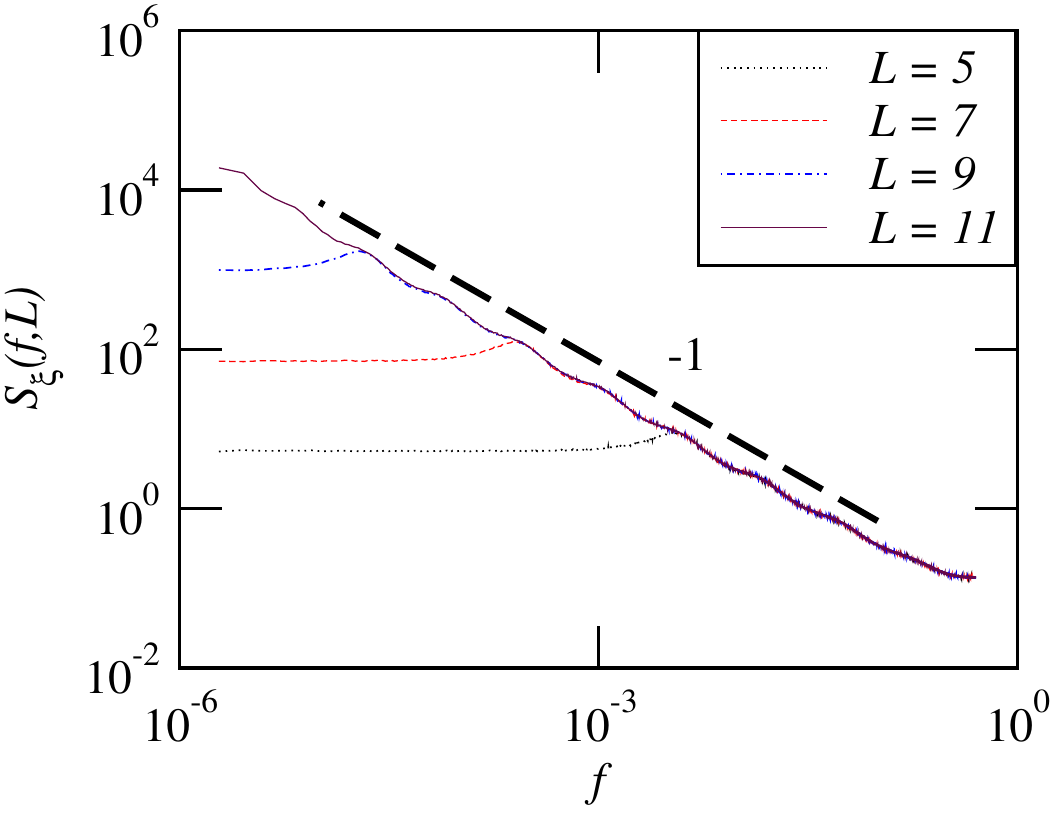}}
  \caption{Power spectra for the fluctuations in the total energy $\xi(t)$  with different system size $L$. Here, $N= 2^{20}$ and $M=10^4$. The dashed line with slope -1 is drawn for comparison. }
  \label{fig_ps_xi}
\end{figure}

\begin{figure}[t]
  \centering
  \scalebox{0.72}{\includegraphics{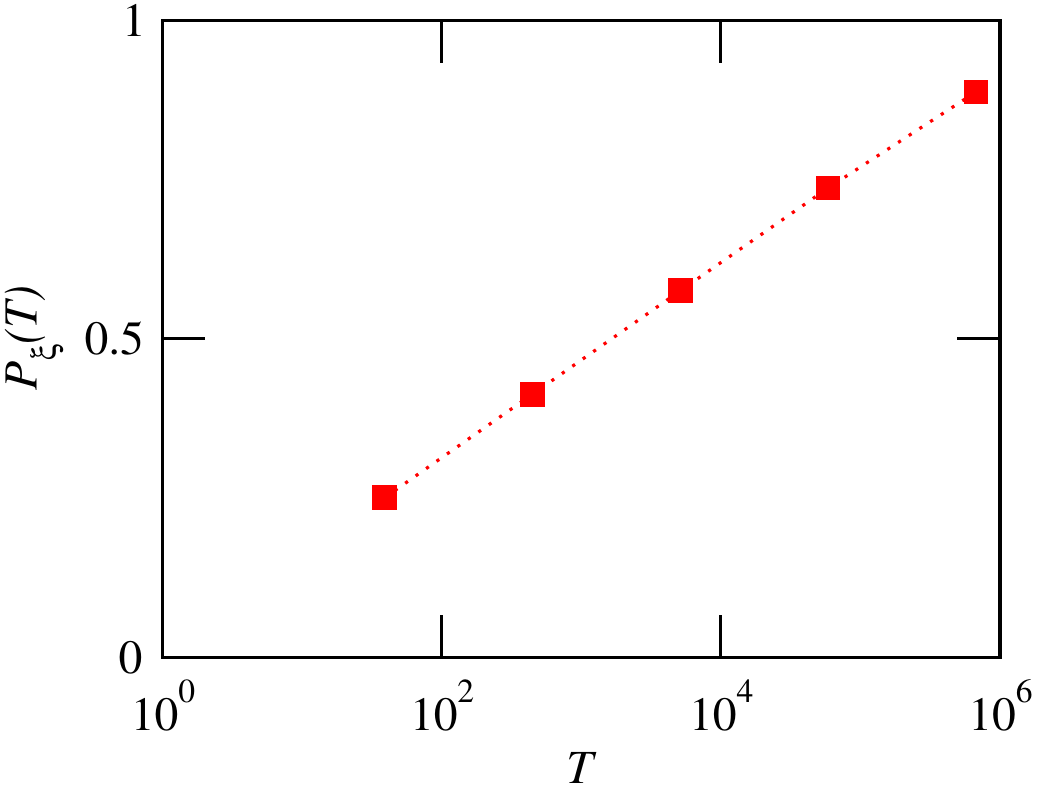}}
  \caption{The variation of the total power $P_{\xi}(T)$ for the fluctuations in the total energy $\xi(t)$ as a function of the cutoff time $T \sim \exp(\mu L)$, with $\mu \approx 1.2$. Clearly, $P_{\xi}(L) \sim L.$}
  \label{fig_pt_xi}
\end{figure}

\subsection{The total energy: $\xi(t)$}
\begin{figure}[t]
  \centering
  \scalebox{0.7}{\includegraphics{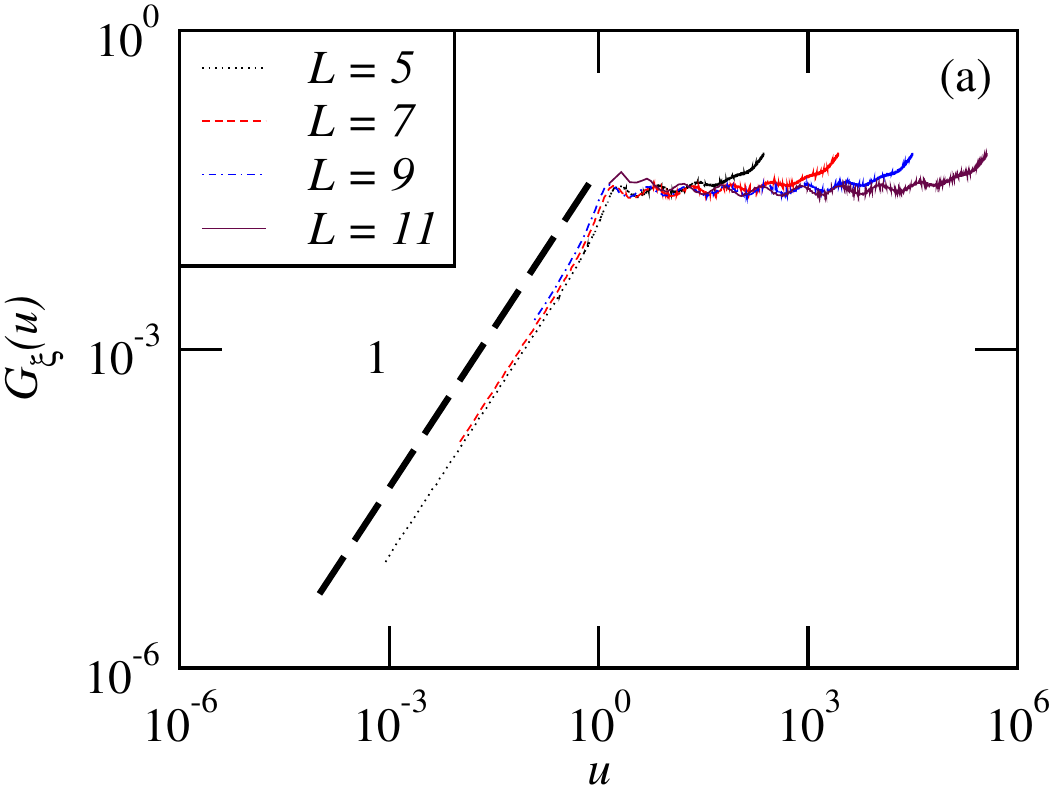}}
   \scalebox{0.7}{\includegraphics{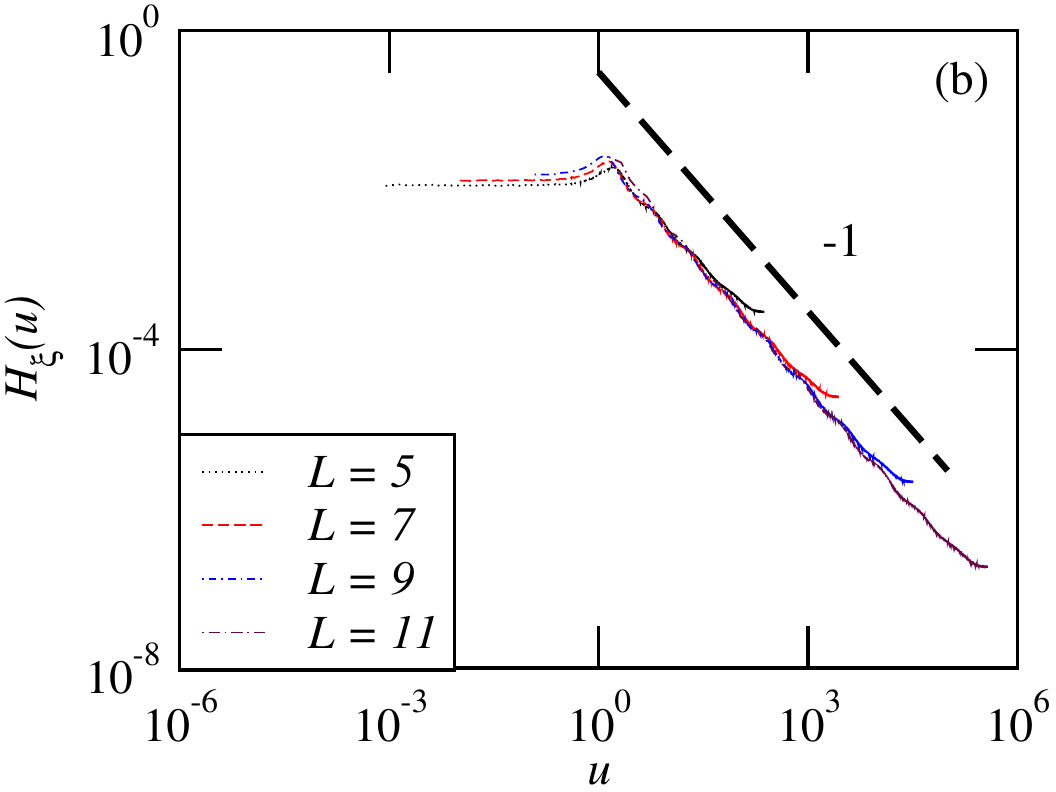}}
  \caption{ The scaling functions. (a) $G_{\xi}(u) \sim fS_{\xi}(f,T)$ with $u = fT$. (b) $H_{\xi}(u) \sim T^{-1}S_{\xi}(f,T)$.}
  \label{fig_xi_scal}
\end{figure}

Figure~\ref{fig_sg_xi} shows a noisy signal $\xi(t)$. We first explore how the system size $L$ affects the spectral behavior of $\xi(t)$. The power spectrum shows $1/f$ behavior for frequencies $1/T\ll f \ll 1/2$, with a cutoff. It also depends on the cutoff time only for $f\ll 1/T$. From the numerical results shown in Fig.~\ref{fig_ps_xi}, it is easy to note that the power spectra follow Eq.~(\ref{a0_psd_eq1}) to Eq.~(\ref{a0_psd_eq3_h}). Interestingly, the cutoff time varies in an exponential manner $T(L) \sim \exp{(\mu L)}$, and 
the total power as a function of the cutoff time varies logarithmically as  
\begin{eqnarray}
P(T) = \int df S_{\xi}(f,T) \sim \int df \frac{1}{f^{1+\epsilon}}G_{\xi}(u)  \sim \ln {T},
\label{eq_pt_xi}
\end{eqnarray}
where $\epsilon \to 0$. Figure~\ref{fig_pt_xi} shows the variation of total power as a function of the cutoff time $T$, and it excellently agrees with Eq.~(\ref{eq_pt_xi}). The exponential behavior of the cutoff time implies that the noisy signal has a large cutoff time even with a small system size. The cutoff behavior in the PSD occurs for $f<1/N$ with $N=2^{20}$ for larger values of the system size ($L> 11$). 
Figure~\ref{fig_xi_scal} shows numerical results for the scaling functions. The data collapse curves satisfy Eqs.~(\ref{a0_psd_eq3_g}) and (\ref{a0_psd_eq3_h}) within the statistical error.

An explanation of the $1/f$ noise is easy to follow based on the scaling analysis of the power spectra for the local energy fluctuations discussed below  [cf. Eq.~(\ref{local_ps_scal})]
\begin{eqnarray}
S_{\xi}(f) = \sum_xS_{z}(f,T_x)\approx \int_{1}^{L} dx S_{z}(f,T_x)  \nonumber \\ \sim \frac{B}{\mu f} \int du \frac{G_z(u)}{u}
  \sim \frac{1}{f},
\end{eqnarray}
where $u = fT_x \sim f\exp(\mu x)$.

\subsection{The local energy: $z(x,t)$}
\begin{figure}[t]
  \centering
  \scalebox{0.62}{\includegraphics{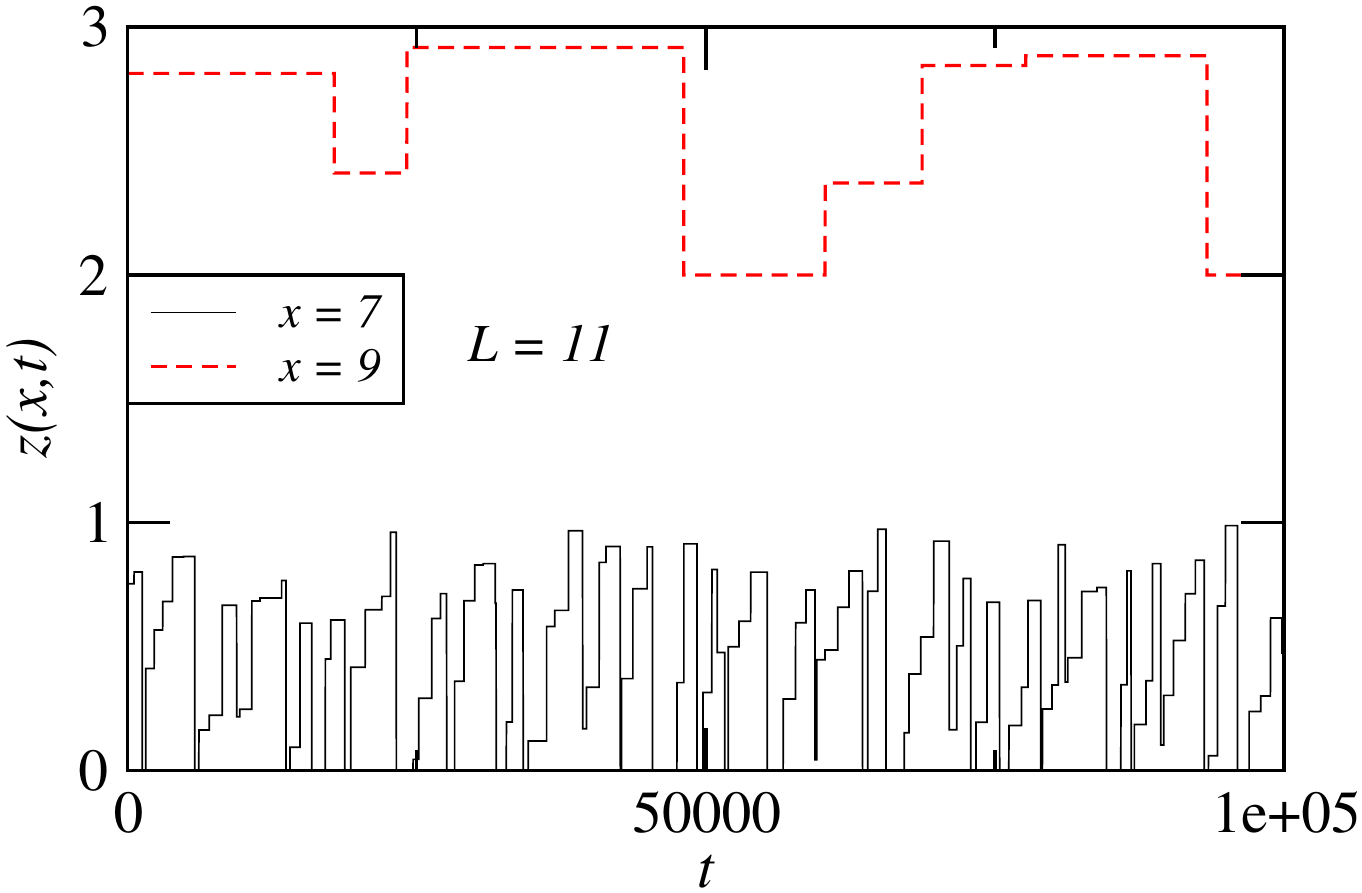}}
  \caption{The local energy $z(x,t)$ as a function of time for two different values of $x$. The system size is $L = 11$, and the upper curve is shifted by 2.}
   \label{fig_sg_z}
\end{figure}

\begin{figure}[t]
  \centering
  \scalebox{0.7}{\includegraphics{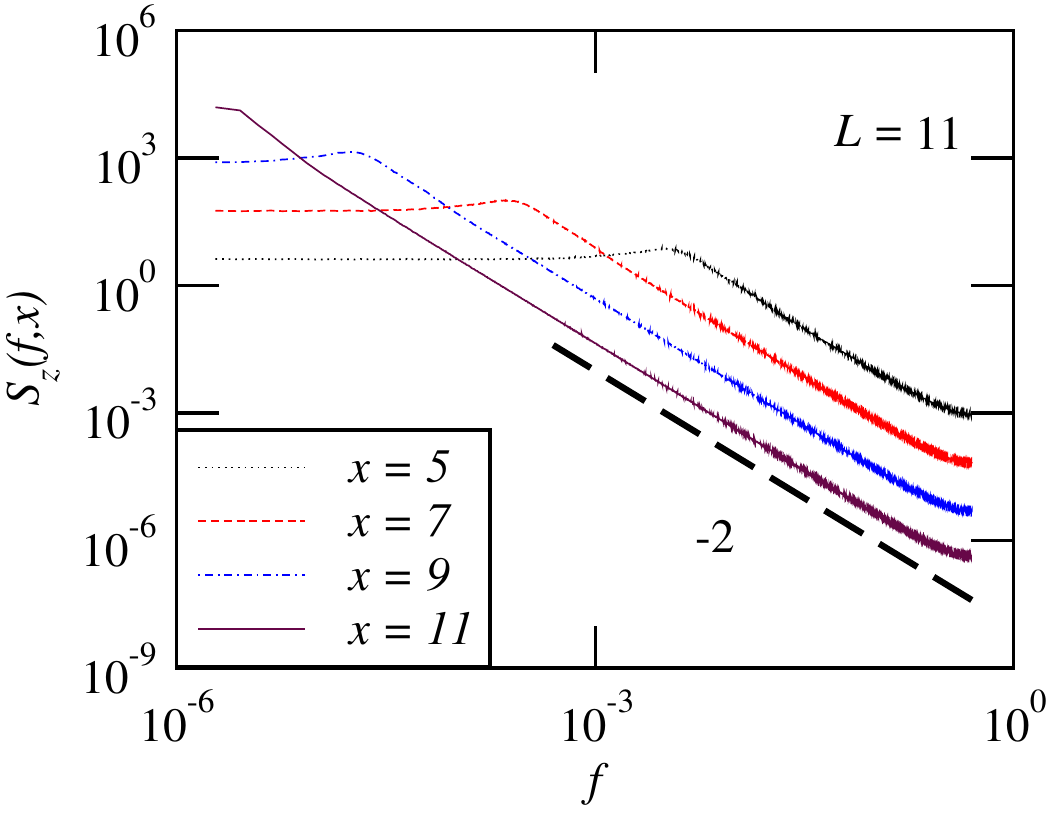}}
  \caption{Power spectra for the local energy fluctuations $z(x,t)$ with different $x$ values. The straight line has a slope -2. }
  \label{fig_ps_z}
\end{figure}

We plot the time variation of $z(x,t)$ for two different values of $x$ in Fig.~\ref{fig_sg_z}. Figure~\ref{fig_ps_z} shows power spectra for the local energy fluctuations $z(x,t)$, suggesting $1/f^2$-type behavior in the frequency regime $1/T_x \ll f \ll 1/2$ with a cutoff. The power explicitly depends on the cutoff time $T_x$ (a function of the distance $x$) for the entire range of frequencies. It is independent of frequency in the regime $f\ll 1/T_x$, but depends on $T_x$ as $S_z(f \approx 1/T_x)$. In the regime $f\gg 1/T_x$, the power value decreases on increasing $T_x$.  One can easily write an expression for the power spectrum
 \begin{equation}
S_z(f,T_x) = \begin{cases} BT^{2-\theta}_{x}, ~~~~~~{\rm for}~~f \ll 1/T_x, \\ B\frac{1}{f^2 }\frac{1}{T^{\theta}_{x}}, ~~~~~~{\rm for}~~1/T_x \ll f \ll 1/2, \end{cases}
\label{local_psd_eq1}
\end{equation}
where $\theta$ is a critical exponent. The cutoff time behaves in an exponential manner $T_x \sim \exp{(\mu x)}$. Recognizing $S_z(f,T_x)$ as a homogeneous function of its arguments [cf. Eq.~(\ref{local_psd_eq1})] and applying the scaling method, we can write
 \begin{equation}
S_z(f,T_x) = B\frac{1}{f^{2-\theta}}G_z(u) = BT^{2-\theta}_{x} H_z(u),
\label{local_ps_scal}
\end{equation}
where $u=fT_x$, and $G_z$ and $H_z$ are scaling functions.

\begin{figure}[t]
  \centering
  \scalebox{0.7}{\includegraphics{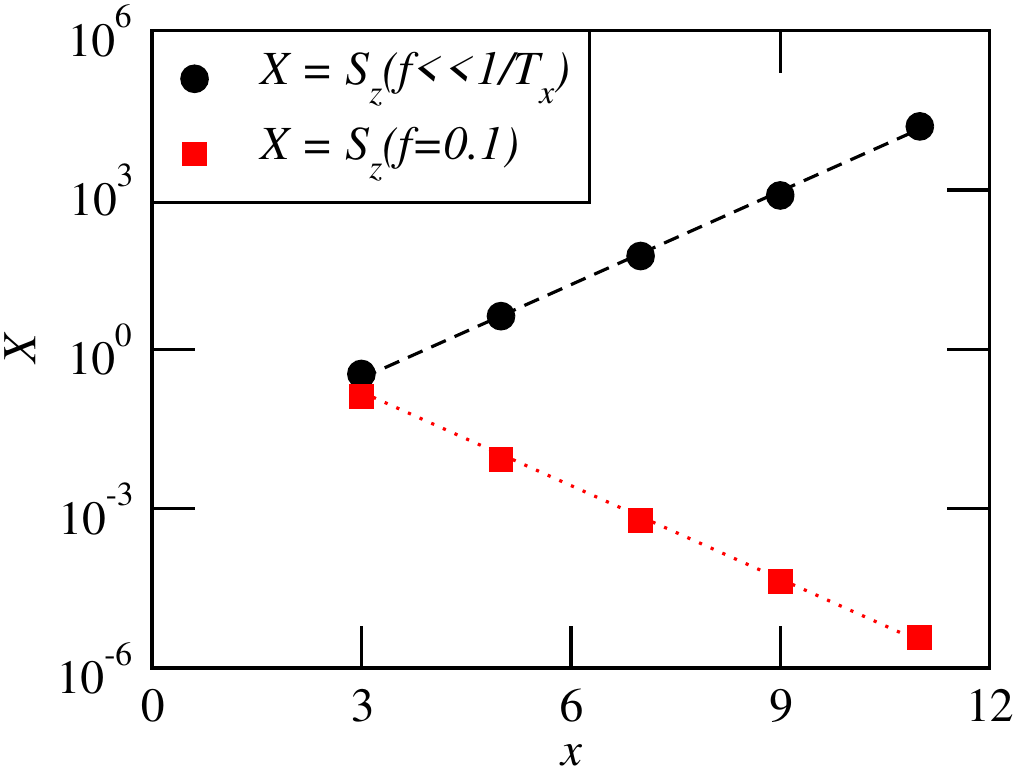}}
  \caption{The power with $x$ in two different frequency regimes, suggesting an exponential behavior $\mu \approx 1.3$. }
  \label{fig_ps_z_1}
\end{figure}

\begin{figure}[t]
  \centering
  \scalebox{0.7}{\includegraphics{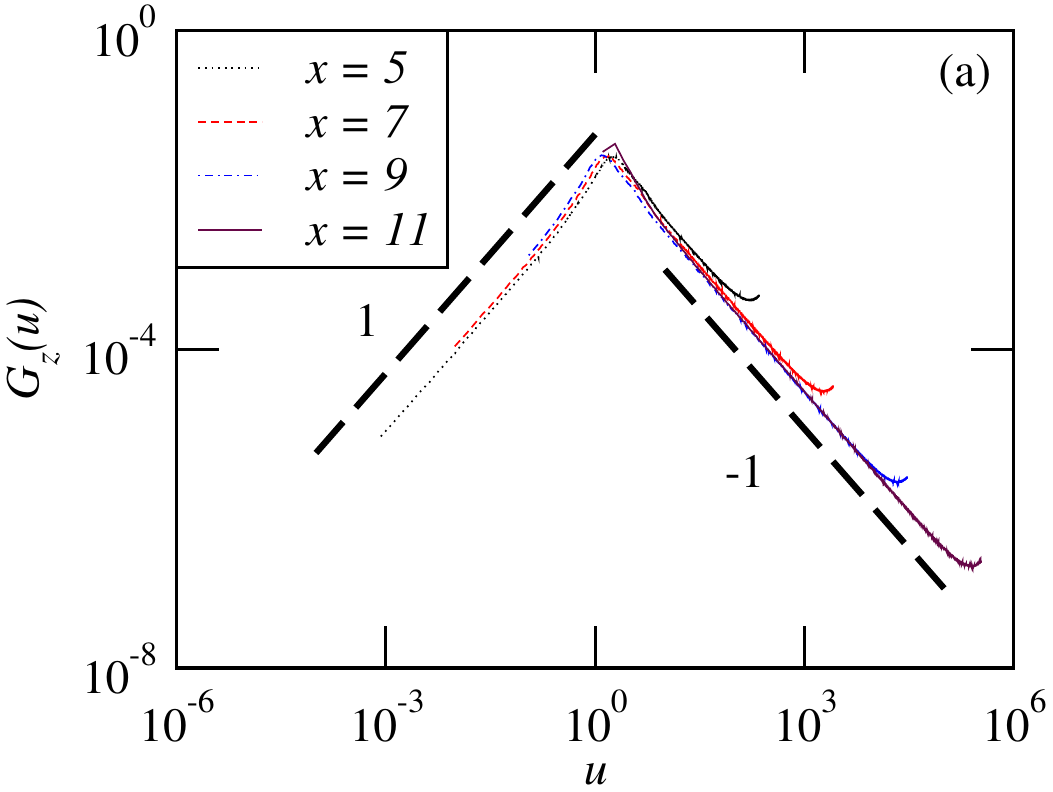}}
   \scalebox{0.7}{\includegraphics{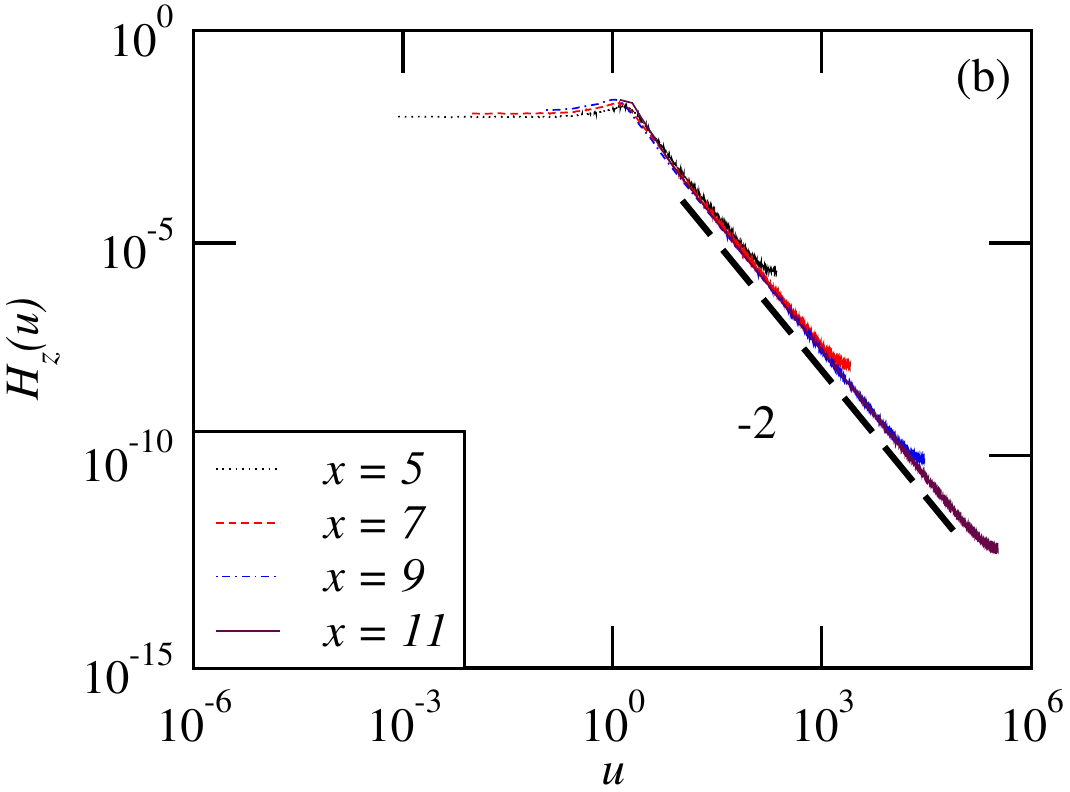}}
  \caption{(a) The scaling function $G_z(u) \sim fS_{z}(f,T_x)$ with $u = fT_x$. (b) $H_z(u) \sim T^{-1}_{x}S_z(f,T_x)$. Here, $T_x \sim \exp(\mu x)$.}
  \label{fig_z_scal}
\end{figure}

The exponent $\theta$ can be numerically estimated, as the total power behaves as
\begin{equation}
P(T_x) = \int df S_z(f,T_x) = B \int df \frac{1}{f^{2-\theta}}G_{z}(u) \sim T^{1-\theta}_{x}.\nonumber
\end{equation}  
The numerical result (not shown) suggests a logarithmic decay, implying  $1-\theta = \epsilon$ with $\epsilon \to 0$. The numerically estimated value of $\epsilon$ is  -0.0007(3).
Then the scaling functions vary as
 \begin{subequations}
 \begin{align}
G_z(u) \sim \begin{cases} u, ~~~~~~{\rm for}~~u \ll 1, \\ 1/u, ~~~{\rm for}~~u\gg 1,\end{cases}
\label{eq_z_g}
\end{align}
and
 \begin{align}
H_z(u) \sim \begin{cases} 1, ~~~~~~{\rm for}~~u \ll 1, \\ 1/u^2, ~~{\rm for}~~u\gg 1.\end{cases}
\label{eq_z_h}
\end{align}
\end{subequations}
Figure~\ref{fig_ps_z_1} confirms the exponential dependence of the cutoff time as a function of the distance from the driving end. Figure \ref{fig_z_scal} shows numerical results for the data collapse curves, excellently consistent with Eqs.~(\ref{eq_z_g}) and (\ref{eq_z_h}). The local energy fluctuations are Lorentzian spectra~\cite{Naveen_2021} with a cutoff time that grows exponentially as a function of distance from the driving end.

\section{Summary}{\label{sec_vi}}
In summary, we have studied a continuous state self-organized critical system, the Zhang sandpile model on one-dimension, with locally conservative or dissipative dynamics. In the critical state, the total energy per site $\xi(t)/L$ hovers about a fixed energy value $ \lim_{L\to \infty}\langle \xi(t)\rangle/L = z_c$, and the energy of a site gets centred about $z_c$ showing a nontrivial peak in its probability density function. For the fixed input noise parameter $\delta_m$, the value of $z_c$ does not depend on the dissipation parameter. One can also estimate the finite spread of the peak from the variance (or integral of the power spectrum over frequency) of $\xi(t)$. We numerically computed power spectral density for the fluctuations in the total energy of the system at the external drive time scale. We observe a trivial behavior for the bulk drive, where the energy fluctuations show Lorentzian spectra with a cutoff time that increases linearly with the system size (cf.~\ref{A1}). While the energy fluctuations show $1/f$-type spectral density for the boundary drive, the cutoff time is a nonlinear function of the system size, varying as power-law or exponentially depending on the local dynamics whether it is conservative or dissipative.

We emphasize that the behavior of cutoff time can reveal the distinct nature of the local dynamics. One implication is that locally dissipative dynamic is not an essential ingredient of the system to observe $1/f$ fluctuations. It is also important to note that the ``directed'' attribute is a necessary condition for such sandpile systems showing $1/f$ noise at the external drive time scale. The $1/f^{\alpha}$ noise observed here seems to occur in systems that have discrete or continuous state and locally conservative or nonconservative dynamics as long as the drive has a preferred direction. However, the spectral exponent may differ from 1. It would be interesting to examine an extent of the $1/f$ noise in a recently introduced discrete state non-conserving sandpile model~\cite{Gros_2020} that mimics the behavior of integrate-and-fire neurons.

\section*{ACKNOWLEDGMENTS}
NK acknowledges the CUJ-UGC fellowship for financial support, and ACY acknowledges seed grants under IOE and a grant ECR/2017/001702 funded by SERB, DST, Government of India.

\appendix

\section{Energy fluctuations under bulk drive}
\begin{figure}[t]
  \centering
  \scalebox{0.7}{\includegraphics{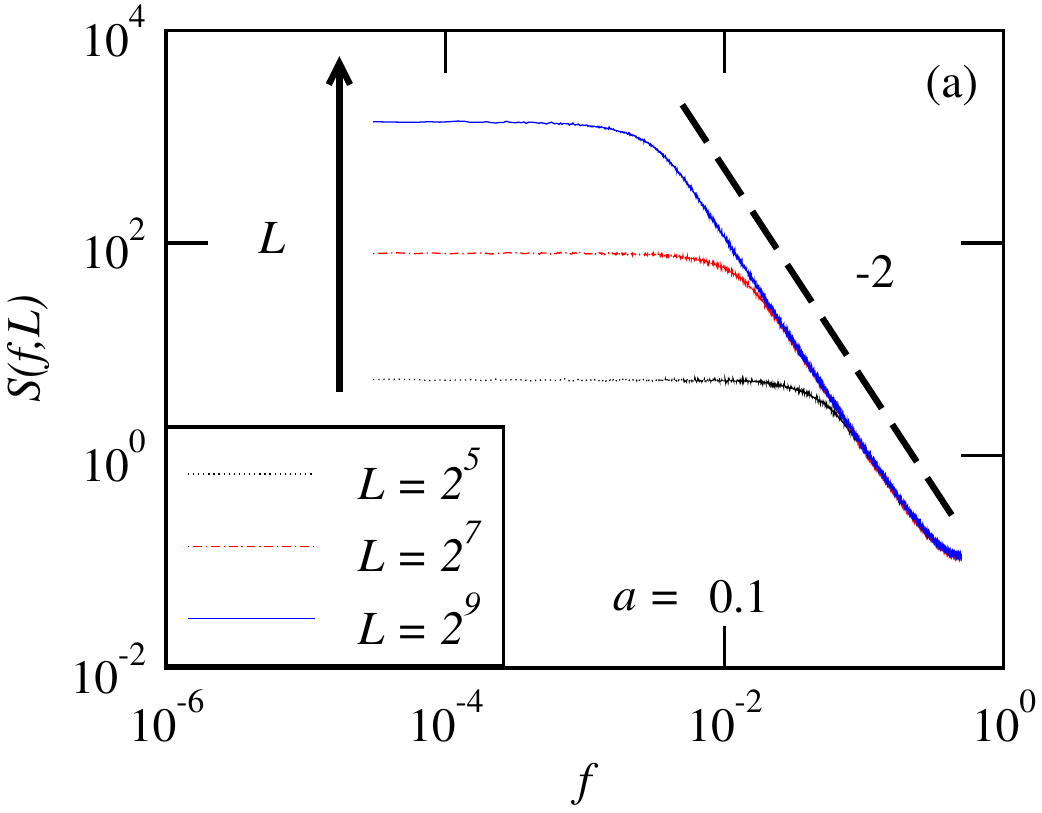}}
   \scalebox{0.7}{\includegraphics{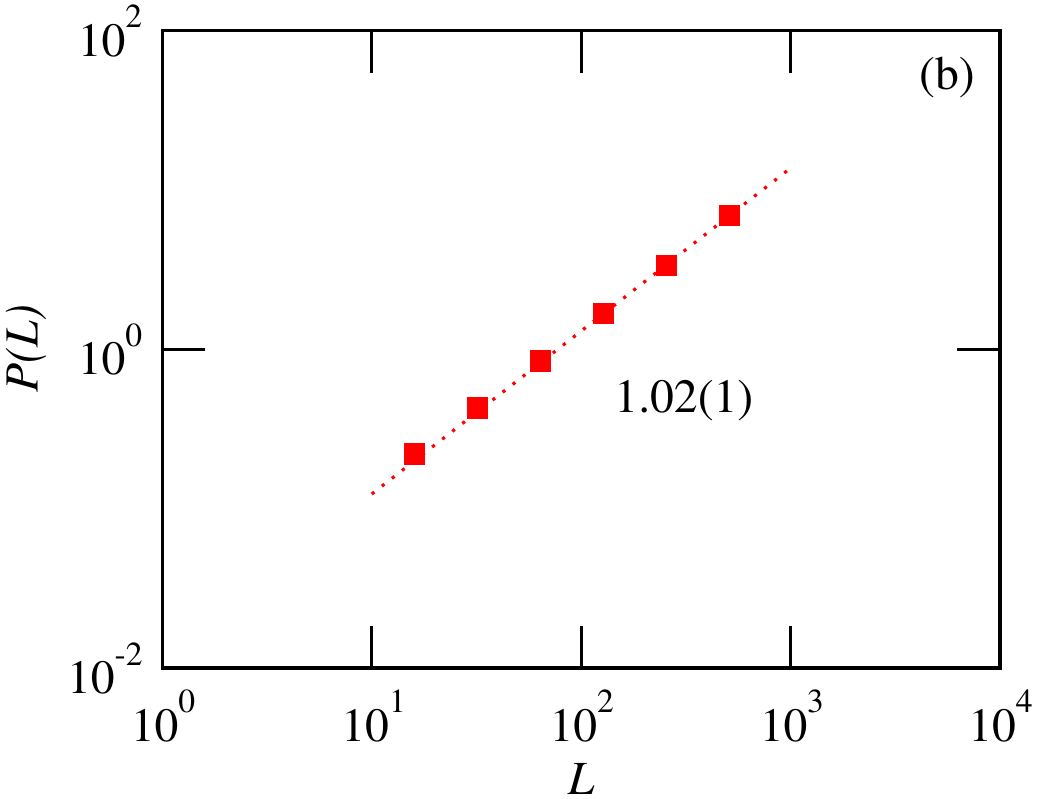}}
  \caption{(a) Power spectra for the fluctuations in the total energy (under bulk drive) for different system size $L$ with $M = 10^{4}$. (b) The total power $P(L) \sim T \sim L$ with the system size. }
  \label{fig_ps_bd_1}
\end{figure}

{\label{A1}}
As shown in Fig.~\ref{fig_ps_bd_1}, the energy fluctuations for nonconservative local dynamics exhibit basically Lorentzian spectra, with a cutoff time that grows in a linear fashion with the system size as $T \sim L$. We also get qualitatively similar results for $a = 0$.

\providecommand{\newblock}{}

\end{document}